# A Policy Report Evaluating the National Assessment Program for Literacy and Numeracy (NAPLAN) Reform in Australia: The Impacts of High Stakes Assessment on Students

Wenya Zhang

Institution of Education, University College London, London, UK

zwenya18@gmail.com

## Abstract

*The National Assessment Program for Literacy and Numeracy (NAPLAN) Reform in Australia, launched in 2008, has emerged as the country's most significant and contentious reform. However, due to its high-stakes nature and standardization, testing presents various challenges. These challenges include the combination of accountability with the 'My School' website, overlooking higher-order cognitive abilities, exacerbating students' anxiety and stress, and creating inequity for Language Background Other Than English (LBOTE) students. This report assesses the achievements and obstacles of the NAPLAN reform, proposing recommendations such as transitioning to online testing, enhancing content and platforms, increasing public assessment literacy, and investing more in LBOTE education. These suggestions aim to strike a balance between standardized testing and authentic educational pursuits, adapting to the evolving needs of students to create a fair, inclusive educational environment that addresses the demands of the 21st century.*

## 1. Introduction

The National Assessment Program for Literacy and Numeracy (NAPLAN) Reform stands as one of the most significant and contested educational reform by the Australian Federal Government over the past two decades [1]. Since its inception in 2008, NAPLAN has garnered extensive attention and has been the focus of critical evaluation. The aim of this report is to analysis of the NAPLAN reform in Australia. This report starts to describe NAPLAN from three perspectives: justification of choosing this reform, year zero and description of this reform. After then, the main body of the report explores the impacts of this high-stakes and standardised testing on students from its introduction in 2008 to the present day. This exploration highlights both the successes achieved and the challenges faced over the years. In response to the findings, the report culminates in a series of considered recommendations. These are: transitioning to NAPLAN's online testing format, refining both the written content and the 'My School' platform, enhancing the public's literacy on assessments, amplifying investments in LBOTE education, and broadening NAPLAN's horizon.

## 2. Description of the Reform

There are two reasons for choosing this reform.

Unique Characteristics and Scale of NAPLAN. NAPLAN is the first and largest national educational test in Australia. Whilst it aligns with international educational trends, it retains characteristics highly specific to the Australian context [2]. Such uniqueness positions NAPLAN as a distinct case of a national experiment in large-scale standardised assessments globally. Examining this distinctiveness offers valuable insights into the successes and challenges of such a pioneering reform in Australia.

The Central Role of Students in the Assessment. This report focuses on the impact and potential benefits of testing on students. Students undoubtedly form the nucleus of any educational assessment; their performance and experiences directly influence the outcomes and efficacy of the Australian National Assessment [3]. It's crucial to note that over a million students participate annually in NAPLAN tests [4]. Furthermore, there's growing interest in academia, evidenced by a growing body of research, that critically analyses the impacts of high-stakes testing (NAPLAN) on students [4,5,6]. The wealth of scholarly discourse on NAPLAN in Australia serves as a robust foundation for this report. This report seeks an understanding of its impact on students and evaluates its alignment with their needs and interests.

## 3. YEAR ZERO

The year zero for this report is 2008– marking the commencement of NAPLAN implementation in Australia. Serving as a vital benchmark in Australia's educational landscape, NAPLAN is engineered to annually assess students in Years 3, 5, 7, and 9 across schools nationwide [2].

This assessment evaluates four key domains: 'reading, writing, language conventions (including spelling, grammar, and punctuation), and numeracy' [7]. These assessing skills are considered quintessential skill for the twenty-first century [8]. Moreover, the results, having undergone standardised testing, present an overall measure of students' literacy and numeracy capabilities, facilitating comparability [8].

Furthermore, NAPLAN has profoundly influenced the educational landscape - from altering educational objectives and reshaping educators' competencies and roles to strengthening parents' engagement in students' learning process [2]. In general, NAPLAN aims to achieve the following five purposes, as outlined by McGaw et al. [7]: '(1) Monitoring national, state and territory programs and policies; (2) System accountability and performance; (3) School improvement; (4) Individual student learning achievement and growth; (5) Information for parents on school and student performance.' (p.9)

## 4. DESCRIPTION OF THE REFORM

NAPLAN employed paper-based testing from 2008 to 2018. This test was developed and centrally administered by the Australian Curriculum, Assessment and Reporting Authority (ACARA). Examination administration was delegated to state and territory authorities [9]. In the evaluation process, students' performances were measured through a blend of digital assessment and expert analysis, all benchmarked against national standards for both summative and formative objectives [7]. These assessments spanned three consecutive days in May, each segment lasting at least 40 minutes [10]. From 2018 onwards, there was a notable shift as more schools adopted the digital NAPLAN test version. By 2023, this transition culminated in the establishment of a new online assessment model emphasizing high-quality teaching and learning, and offers richer national assessment information [8].

A consequential development in NAPLAN was the introduction of the 'My School' website in January 2010, which imparted a pronounced high-stakes characterisation to the assessment [5,11]. Such high-stakes assessment (HSA) stands out in educational reforms. Both public and private educational institutions across states unveil, collate, and present an annual compilation of students' performance—NAPLAN's literacy and numeracy performance—alongside demographic and fiscal data on this federal platform [11,12]. The primary purpose of the website is to enable the public to compare schools catering to students with similar socio-educational backgrounds [7,11]. In the context of schools, it has evolved into an indispensable instrument underpinning accountability [5]. However, this accountability sees an increase in stakes, exacerbated by media coverage. Often sidelining ACARA's original vision—which advocates against ranking schools based on sheer outcomes without accounting for disparities in student backgrounds—the media leverages these results to spotlight and, at times, impose pressure on schools perceived as lagging [13,7]. It is also likely that some parents who misuse rankings will keep their children away from schools that did not perform well on NAPLAN [10]. Consequently, NAPLAN became the HSA from 2010 onwards [5,7].

NAPLAN reform is also notable for its function as national standardized test. Its start dates back to 2008; however, over the preceding decade, several Australian states and regions had already employed standardised testing within their educational frameworks [11]. Notably, these state-specific assessments differed considerably in their format and were governed by distinct regulations [12]. The landscape began to shift in 2007 when the Rudd/Gillard Labor administration incorporated NAPLAN into a broader school reform strategy, a move that secured bipartisan support from the outset [2]. The subsequent year witnessed the signing of two foundational documents: the Melbourne Declaration on the Educational Goals for Young Australians and Quality Education: The Case for an Education Revolution in Our Schools. Both collectively entrenched NAPLAN within the central architecture of governmental educational reform [11]. By collecting standardized data and promoting international standards for curriculum and testing [14], the federal government hopes to improve student achievement [15]. Through this reform, they intend to improve Australia's international trends in literacy and numeracy [16]. In summary, NAPLAN catalysed the conception and crystallisation of a standardised and national educational system [17].

## 5. SUCCESSES AND CHALLENGES OF THIS REFORM

NAPLAN's educational reform emerges from a multifaceted international assessment environment [18,2]. This national test mirrors the culture of accountability and high-stakes testing that has long been ingrained in the education systems of the United States and the United Kingdom [10]; It is also moulded by the overarching international neoliberal policy framework, which champions the privatisation of education to augment school effectiveness and elevate educational standards [19]; Concurrently, NAPLAN aligns with the tenets of the Global Education Reform Movement (GERM) that has found traction in Australia [2]. GERM prioritises 'top-down, test-based modes of accountability interlinked with parental choice and market reforms, encompassing bureaucratic restructuring and diverse forms of privatisation' [17]. However, in international assessments such as the Programme for International Student Assessment (PISA) and the Trends in International Mathematics and Science Study (TIMSS), Australia consistently ranks in the mid-tier. It trails behind several Asian countries and often lags behind nations like England and Canada [7]. This landscape has imbued NAPLAN with an implicit task: to assuage national concerns stemming from subpar performances in global standardised tests [20]. Hence, NAPLAN's inception can be viewed as a response to the evolving demands of the complex international assessment environment.

A review of these trends in the international assessment environment points to two terms: one is high stakes assessment (HSA) and the other is standardised assessment (SA). This is the distinctive feature of NAPLAN reform. Therefore, the next section will delineate the successes and challenges of these features.

### 5.1. Successes

NAPLAN implementation has marked several accomplishments in the Australian educational landscape. In the context of standardised testing, NAPLAN has emerged as an effective tool. It enables stakeholders-students, parents, and educators-to assess students' aptitudes for literacy and numeracy. This has facilitated targeted interventions where necessary, driving marked progress in student performance over the years. NAPLAN has thus efficiently met its two core purposes: to inform and improve student achievement.

### 5.1.1. Informing Teachers and Parents of Student's Performance

One of the standout benefits of NAPLAN is its instrumental role in delivering invaluable insights into student performance for teachers, parents, and educational institutions. The primary purpose of the NAPLAN test is to determine whether students have literacy and numeracy skills [8]. Utilizing the results of the standardized test, educators and institutions combine NAPLAN outcomes with complementary datasets to diagnose areas where students may be lagging or falling short of benchmarks [21]. This methodology illuminates pathways for betterment, empowering educators to provide timely and impactful support to students [21]. For instance, in the context of high stakes and standardised assessment, professional teachers are familiar with the content of standardised tests and actively seek out test preparation materials to bridge performance gaps and enhance student outcomes [4].

On the parent engagement front, Gillard proclaimed that 'A new era of transparency - My School - has put information in the hands of parents and created a new national conversation about what we should expect from our schools and about what it takes to deliver great results' [14]. This transparency granted parents the right to publicly access NAPLAN results online regarding their children's performance [22]. The 'My School' website underwent a significant overhaul in 2020, allowing for comparisons of student performance both across different schools and longitudinally within the same school over multiple years [23]. Mills asserts that these personalized reports provide parents with a more comprehensive understanding of their children's academic progress [24]. The 'My School' platform further empowers parents to draw parallels between schools, providing a comparative lens grounded in analogous educational contexts, pertinent to their children's academic voyage [7]. Such insights significantly shape parents' choices, guiding their decisions when selecting the most fitting educational settings for their offspring [10].

To sum up, NAPLAN's invaluable feedback mechanism not only equips schools and parents with crucial data on student performance in literacy and numeracy but also facilitates the implementation of targeted instructional approaches [7].

### 5.1.2. Improvement in Student Performance

One of the significant accomplishments attributed to the NAPLAN reform is the measurable enhancement in student performance. Thompson and Harbaugh claim that improving student performance in the NAPLAN reform program is essential to the country's competitiveness in the international economy [15]. Sloane and Kelly suggest that HSA can motivate students to achieve better academically [25]. This belief is rooted in the idea that such assessments offer students a transparent lens into their skills and knowledge and a clearer understanding of what is emphasized in the course, which motivates them [21].

Empirical evidence supported this hypothesis. The 2020 Australian Federal Government's review of NAPLAN reported an uptrend in reading and numeracy scores among primary school students [7]. Further bolstering this narrative is the increased proportion of Australian students in grade 4 marking their prowess on international platforms such as PISA (for reading) and TIMSS (for mathematics). Delving into age-specific performance, Australia's 15-year-olds have consistently outperformed the global mean in PISA scores across reading, mathematics, and science over the past decade [26]. While the Organisation for Economic Cooperation and Development (OECD) published an Economic Survey for Australia and issued a caution that a slight regression in the nation's PISA scores vis-à-vis its global counterparts in recent years [26], notable improvements in performance for primary and 15-year-old cohorts cannot be denied. Therefore, in this regard, the NAPLAN reform has enhanced student achievement.

However, in the more than 20 years since NAPLAN was implemented, more and more critics have claimed that the test has had unexpected negative consequences. As stated by Polesel et al. [27], high stakes testing 'like fire, is a wonderful servant, but a very poor master.' (p.653).

### 5.2. Challenges

A report of NAPLAN in 2020 revealed concern about the distortion of education and practice caused by NAPLAN, the misuse and misunderstanding of the data of the 'My School' website [7]. Further, some studies have shown that it also exacerbates students' stress and anxiousness, and unfair treatment of LBOTE students in Australia.

#### 5.2.1. Limitations of Measuring Student's Performance

NAPLAN has the advantage of being highly comparable because of its standardized nature [8]. However, this nature of standardization has resulted in measured errors for individuals. According to Wu [28], standardised tests (NAPLAN) are often inaccurate as a measure of school and student performance due to their wide margin of error. A similar finding was made by Ladwig [29], illuminating that a significant 12% of students experience issues with measurement errors in their NAPLAN outcomes. The cause of the error may be the lengthy feedback loop, with a period over five months between test-taking and feedback reception [30]. Harris et al. [31] encapsulate this dilemma, suggesting that while NAPLAN might be apt for macro-level evaluations, it remains an imperfect tool for individual performance assessments.

The construct of certain sections of the test, like the writing segment, has come under scrutiny, with critiques challenging its validity. There are some studies that have criticized the form of the writing test design, administration and reporting of the NAPLAN test, arguing that this part lacks validity [7]. In other words, the NAPLAN reading test results do not prove that students write excellent articles [32,7].

#### 5.2.2. Equity Concerns for LBOTE Students

Equity, a cornerstone of the NAPLAN ethos, has ironically emerged as one of its profound challenges. While NAPLAN aspired to be the panacea for academic disparities, ensuring equitable resource distribution [8,14], in practice, it has exacerbated the gap between different student cohorts [33].

A glaring oversight in the NAPLAN framework is its treatment of the LBOTE (Language Background Other Than English) demographic. NAPLAN recognises diversity by classifying students as Indigenous or LBOTE for testing [2]. The inherent design bias, favouring standard Australian English, inadvertently alienates a mixed cohort of native speakers, ESL (English as a Second Language) learners, and EFL (English as a Foreign Language) learners, especially those coming from remote Indigenous communities [34]. This linguistic and cultural bias, rather than bridging gaps, accentuates disparities. A government report confirms this inequity, shedding light on the subpar performance of these demographics, which is even more pronounced among those from remote rural settings [34].

#### 5.2.3. Misinterpreted My School's Data and Misguided Decision Making

The transparency introduced by the 'My School' website was initially envisioned as a tool to aid parents in making informed decisions (see section 5.1 above). However, its implementation has resulted in unintended consequences. A major factor contributing to this issue is the misinterpretation of data, not only by the general public but also, alarmingly, by media institutions.

Parents may misinterpret the data [28,29]. The complexity of data representations on the 'My School' website necessitates a nuanced understanding of statistics for accurate interpretation. However, as highlighted by Rose et al. [2], navigating these charts and figures requires a specific proficiency in statistical interpretation or experts with corresponding assessment literacy. Consequently, not all stakeholders, especially parents, possess the necessary statistical understanding to fully grasp the subtleties of the results [17].

This misinterpretation of data is exacerbated by the media's penchant for sensationalism. Since many parents have limited statistical literacy, they often turn to league tables presented by Australian media outlets [35]. These rankings are frequently mischaracterized as definitive measures of school quality [2]. However, the information contained within these rankings is notably limited, as they primarily compare students' literacy and numeracy performance within these institutions [7]. By fixating on these data, parents may overlook schools that provide a diverse range of educational experiences and resources beyond mere standardized test scores [35].

Therefore, this misinterpretation has inadvertently encouraged the creation of league tables by media outlets, distorting perceptions of school quality and potentially misguiding parental choices for their children's education [5,11,36,].

### 5.2.4. Elevated Student Anxiety and Pressures from Testing

Furthermore, public reporting on 'My School' also exacerbated parents' and students' concerns about potential public 'naming and shaming' (p.84), which increased their anxiety [37].

Extensive research in Australia highlighted the negative effects of HSA, particularly anxiety and stress among students. Rogers et al. [6] conducted a survey across 11 independent schools in Western Australia. This revealed that students faced considerable stress during NAPLAN testing. These findings resonate with another study, which indicated that students experienced strong negative emotions, including fear, anxiety, and sadness, throughout the NAPLAN preparation and testing phases [9]. Such feelings are notably intense among older students, with Year 7 students exhibiting increased apprehension due to concerns over NAPLAN's potential implications for their academic and tertiary education access [38]. Furthermore, chronic stress is a known precursor to various health problems [39]. For instance, Athanasou [40] found that examination-related stress among Australian primary school students was linked to health issues like insomnia. However, a contrasting perspective occurs in environments where educators consciously minimize the emphasis on NAPLAN, resulting in diminished negative feelings among students [41]. Consequently, addressing the increasing stress and anxiety associated with NAPLAN testing is of paramount importance.

### 5.2.5. Narrowing the Curriculum and Ignoring Students' Advanced Abilities

High-stakes and standardized tests may distort teaching practices, leading educators to prioritize testing proficiency. Thompson's [42] research conducted with 961 teachers across two Australian states uncovered that NAPLAN reduced time and emphasis on subjects beyond literacy and numeracy. This is attributable to the perception of these tests as high-stakes, with students' grades directly tied to their performance, thereby imposing considerable pressure on educators [42]. Consequently, within this HSA context, there's an underlying expectation that an intensifying focus on test-related content will correlate with improved student scores [43,44]. This narrowed pedagogical approach restricts the curriculum breadth, pushing educators to teach to the test in anticipation of superior student performance [6].

In addition, the NAPLAN framework overlooks students' advanced abilities. The Queensland Studies Authority (QSA) [45] expressed concerns that test-taking encourages 'methods of teaching that promote shallow and superficial learning rather than deep conceptual understanding and the kinds of complex knowledge and skills needed in modern, information-based societies' (p. 21). In accordance with this, Swain et al. [9] advocated that high quality assessments should reflect the richness of the curriculum such as 'critical, creative and higher order thinking' (p.318). However, in the face of HSA, educators often fall into the trap of emphasizing test-aligned content and bypassing higher-order thinking skills [43]. It is paramount to recognize that for students – our future citizens – these advanced cognitive competencies are indispensable tools to navigate an intricate, dynamic world [46,47]. Therefore, NAPLAN should not neglect the development of other subjects and higher-level thinking skills suitable for 21st-century students.

## 6. CONCLUSION

NAPLAN's introduction has indisputably reshaped the educational framework in Australia over the past two decades. Its pivotal role, as highlighted by McGaw et al. [7], is in providing invaluable data on student literacy and numeracy performance. Beyond mere assessment, it serves as a diagnostic tool, offering intervention and support mechanisms for educators, institutions, and even government bodies [50].

However, the challenges of HSA, emphasized by public reporting and accountability [51], have emerged. The accountability for school performance, 'My School' and media propaganda, has inadvertently increased high stakes. Concerns regarding questions about the test's validity, equity, and the narrowing of the curriculum. These concerns hint at the potential destruction of holistic education, sidelining higher-order cognitive abilities and imposing test pressures on students.

Based on these challenges, this report recommends the implementation of effective measures to reduce hazards in HSA. From the perspective of students, returning to the original intention of NAPLAN reform, by balancing standardization with genuine educational pursuits. The revised NAPLAN should be tailored to students' evolving needs and interests.

Subsequently, this report advocates for a suite of recommendations: transitioning to NAPLAN's online testing, refining the written content and the 'My School' platform, enhancing public assessment literacy, bolstering investment in LBOTE education, and broadening NAPLAN's scope.

NAPLAN, as a cornerstone of Australian educational reform [1], demands ongoing reflection and refinement. High stakes assessment should not [52] 'tail starts to wage education dog' (p.4). High-stakes testing, akin to tinder held by Prometheus, should be applied with discernment and caution. It's imperative for all stakeholders to harness its potential responsibly, ensuring it illuminates the path of Australian educational reforms. Creating an educational environment that is equitable, holistic, and capable of meeting 21st-century challenges for Australian students.

## 7. RECOMMENDATIONS

NAPLAN's standardized assessment data has become an instrumental tool over the past two decades, providing insights into student progress and achievement at national, regional, and local levels [7]. Consistent with other Australian studies [2,7], this report acknowledges NAPLAN's role in enhancing student performance assessment and public reporting. Nevertheless, it also highlights concerns regarding its validity in measuring student capability, equity for LOBTE students, implications of 'My School' data utilization, heightened test-induced student anxiety, curriculum restrictions, and overlooking advanced cognitive skills. Consequently, this section delves into these challenges and offers recommendations aimed at aligning the assessment more closely with students' rights and best interests.

## 7.1. Use of NAPLAN Online Test Countrywide and Content Redesign for Writing

Addressing concerns from 5.2.1, this report suggests enhancing feedback timeliness by accelerating result dissemination. Computerized marking improves efficiency over traditional paper-based testing. The dynamic and alive nature of digital processing allows data to be returned to schools within days, compared to the 5-month duration of paper tests [30]. Additionally, a redesign of the writing test is proposed. Incorporating teacher formative assessments into the NAPLAN writing summative assessment could be beneficial [48]. However, online testing brings forth challenges concerning construct validity, necessitating further research to ensure NAPLAN's validity.

## 7.2. Increasing investment in LBOTE education

Reflecting on 5.2.2, teacher training should be intensified, particularly in ESL and EFL for children in linguistically diverse regions. Schools with a significant number of LBOTE students should receive specialized programs tailored to their needs. By receiving focused instruction in standard English, these students are better equipped for NAPLAN assessments [12].

## 7.3. 'My School' Website Redesign

Considering 5.2.3, this report recommends updating the 'My School' website to minimize potential drawbacks. The 'NAPLAN explained' section should be made parent-friendly, ensuring simplicity and accessibility. Ranking authority should be vested in educational entities rather than media outlets. Rankings should reflect an array of factors, encompassing academic performance, teaching staff quality, and available educational resources. For methodology insights, globally renowned university ranking systems like Times Higher Education World University Rankings (THE) and Quacquarelli Symonds World University Rankings (QS) could serve as benchmarks.

## 7.4. Boosting Public Assessment Literacy

To alleviate test-related anxieties (see section 5.2.4), it is crucial to enhance public understanding of assessment procedures. Students should be encouraged to see NAPLAN not just as a performance indicator, but as a developmental tool designed to highlight areas for growth and improvement. The results of NAPLAN are not intended to affect students' future academic prospects directly. By clearly communicating this to students and their families, schools may help shift perceptions and reduce the stress associated with these assessments.

Schools should take the lead in promoting public assessment literacy. This could be achieved by providing clear explanations of NAPLAN's objectives and processes to both parents and students before and after the assessments. Effective communication of test results is also critical—schools need to ensure that parents fully understand their child's performance in context, without misinterpreting the results as absolute measures of success or failure. Workshops, seminars, and simplified reports may help bridge the gap between schools and homes, fostering a deeper understanding of the purpose of NAPLAN.

Beyond that, public assessment literacy requires more than just understanding the test results — it means also being able to understand how these assessments are part of a broader picture about evaluating schools or teachers. Although NAPLAN scores can be influential in assessing schools and teachers, they should not only way to evaluate. Schools and educational authorities need to find a way of mitigating the emphasis on NAPLAN so that students are not put in an unnecessarily stressful environment [41].

### 7.5. Broadening NAPLAN's Horizon

Reflecting on section 5.2.5, NAPLAN should become more comprehensive by integrating subjects such as natural science, mirroring international assessments like PISA and TIMSS. Observing educational strategies from high-performing countries on these tests, such as Singapore [49], could be insightful.

Additionally, the Australian curriculum should be more inclusive, emphasizing real-world applications and fostering creativity, critical thinking, and problem-solving skills, thereby aligning student capabilities with 21st-century societal demands.

In this context, educational institutions should be granted enhanced autonomy and foster an environment conducive to innovation. Such empowerment would position school administrators to pioneer novel approaches in curriculum design, diversify the range of courses available, and assert greater authority over their respective establishments [49]. By adopting such strategies, students could adapt more effectively to PISA and TIMSS assessments and potentially improve their performance.

Initially, pilot tests can be conducted in select regions, encompassing not just numeracy, literacy, and natural science, but also real-world applications and skills such as creativity, critical thinking, and problem-solving. After refining the process, these comprehensive elements can be incorporated into the national testing framework, aligning with the broader objectives of enhancing student capabilities and preparing them for 21st-century challenges.

## ACKNOWLEDGEMENTS

The author is grateful to Professor Mary Richardson for her insightful suggestions on the initial draft of this work during the author's postgraduate studies at UCL.

**Authors**

Wenya Zhang, an alumna of UCL, holds a Master's in Education (Assessment) with a focus on educational assessment and policy.

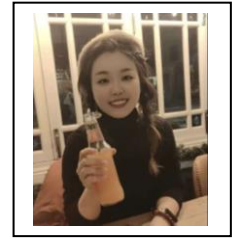